# Reducing T-count and T-depth in approximate quantum Fourier transform circuits

Byeongyong Park[1,2] and Doyeol Ahn[1,2,*]

[1]*Department of Electrical and Computer Engineering and Center for Quantum Information Processing, University of Seoul, 163 Seoulsiripdae-ro, Dongdaemun-gu, Seoul 02504, Republic of Korea*

[2]*Singularity Quantum Inc, 9506 Villa Isle Drive, Villa Park, CA 92861, USA*

**Corresponding author: dahn@uos.ac.kr*

**Abstract**

The quantum Fourier transform (QFT) is a fundamental component in various quantum algorithms, including Shor's factoring algorithm and the Harrow-Hassidim-Lloyd (HHL) algorithm for solving systems of linear equations. Efficient implementation of the QFT is crucial for the practical realization of large-scale quantum algorithms, especially in fault-tolerant quantum computing. In fault-tolerant implementations, the Clifford + T gate library is the standard choice for building quantum circuits, with the T gate being the most resource-intensive component. This paper presents two novel $n$-qubit approximate quantum Fourier transform (AQFT) circuits with an approximation error of $O(\varepsilon)$: AQFT Circuit 1 and AQFT Circuit 2. AQFT Circuit 1 is designed to minimize the number of T gates (T-count). It achieves a T-count of $4n\log_2(n/\varepsilon) - O(\log^2(n/\varepsilon))$ which is roughly half of the best-known result, $8n\log_2(n/\varepsilon) - O(\log^2(n/\varepsilon))$. This optimization is accomplished by constructing inverse phase gradient transformation (PGT) circuits in the QFT circuit without using additional non-Clifford gates. AQFT Circuit 2 focuses on reducing the depth of T gates (T-depth), decreasing it from $O(n\log(n/\varepsilon))$ to $O(n\log(\log(n/\varepsilon)))$. This reduction is achieved by employing logarithmic-depth quantum adders for implementing inverse PGTs. Additionally, to further decrease the T-depth, inverse PGTs are paired and parallelized using only additional $n + O(1)$ T gates. These reductions in T-count and T-depth facilitate the fault-tolerant implementation of the QFT, potentially accelerating the realization of quantum algorithms in practical applications.

## I. INTRODUCTION

The quantum Fourier transform (QFT) is considered one of the most versatile components in quantum algorithms. It plays a key role in fundamental quantum algorithms such as quantum phase estimation [1], algorithms for hidden subgroup problems including Shor's factoring algorithm [2], the Harrow-Hassidim-Lloyd algorithm (HHL) algorithm for solving systems of linear equations [3], and quantum amplitude estimation [4], to name a few. The applications of QFT span various fields, including basic arithmetic operations [5−8], cryptography [9−14], signal processing [15, 16], quantum simulation [17−20], quantum machine learning [21−25], quantum finance [26−28], and computational fluid dynamics [29−31]. Improving the efficiency of QFT implementations is therefore crucial for enhancing the performance of various quantum algorithms and expanding their applicability.

Large-scale quantum algorithms, such as Shor's factoring algorithm [2] and the HHL algorithm [3], which utilize QFT, should be implemented in a fault-tolerant manner due to the fragility of quantum information. In fault-tolerant quantum computations, circuits are synthesized using universal and fault-tolerant gates. The Clifford + T gate library is generally chosen for this purpose in various promising error correction codes [32, 33]. Within these fault-tolerance approaches, Clifford gates are easier to implement, often transversally. In contrast, T gates require more resource-intensive methods, such as magic state distillation [34, 35], thus dominating the implementation cost [36, 37]. Therefore, reducing the number of T gates (T-count) and the depth of T gates (T-depth) is essential for efficiently executing large-scale quantum algorithms.

In fault-tolerant regimes, QFT is approximately implemented with an acceptable error, known as approximate quantum Fourier transform (AQFT). Typically, AQFT is implemented by removing all controlled rotation gates with angles below a certain threshold value. It has been demonstrated that, for effectively implementing various quantum algorithms utilizing QFT, applying AQFT instead of a full QFT often yields satisfactory results without significant performance penalties [38−41]. For instance, using AQFT with a threshold value of $\pi/2^8$ is sufficient for factoring RSA-2048 with a 95% success rate [41].

The basic implementation method for an $n$-qubit AQFT involves approximating the $n$-qubit QFT by omitting small-angle controlled rotations and decomposing the remaining controlled rotation gates into the Clifford + T gate library. This approach reduces the T-count for QFT implementation from $O(n^2 \log n)$ to $O(n \log^2 n)$, while omitting the dependence on the approximation error for brevity. For the semiclassical version of the AQFT [42], which is followed by measurement, thereby limiting its application in the midst of computation, it has been demonstrated that AQFT can be implemented with a T-count of $O(n \log n)$ [43]. In the case of fully coherent AQFT, Nam et al. [44] achieved a T-count of $O(n \log n)$ using a method originally reported in Ref. [45] that involves implementing phase gradient transform (PGT) with a quantum adder. They utilized Toffoli gates (specifically, relative phase Toffoli gates and measurements) to construct PGT circuits and employed quantum adders from Ref. [46]. The T-count and T-depth of their $n$-qubit AQFT circuit with an approximation error of $O(\varepsilon)$ are reported as $8n \log_2(n/\varepsilon) - O(\log^2(n/\varepsilon))$ and $n \log_2(n/\varepsilon) + O(n)$, respectively.

In this paper, we present two new fully coherent $n$-qubit AQFT circuits with an approximation error of $O(\varepsilon)$: AQFT Circuit 1 and AQFT Circuit 2. AQFT Circuit 1 is designed to reduce the T-count. In this design, we construct the rotation gate layers that perform inverse PGTs without using Toffoli gates, which are non-Clifford gates and accounted for approximately half of the T-count in the AQFT circuit design in Ref. [44]. Next, we replace the rotation gate layers with quantum adders from Ref. [46]. On the other hand, AQFT Circuit 2 is designed to reduce the T-depth. In this design, we construct the rotation gate layers without Toffoli gates, pair and parallelize two rotation gate layers into one, and replace them with logarithmic-depth quantum adders from Ref. [47].

As a result, AQFT Circuit 1 achieves a T-count of $4n \log_2(n/\varepsilon) - O(\log^2(n/\varepsilon))$ with a T-depth of $n \log_2(n/\varepsilon) + O(n)$, and AQFT Circuit 2 achieves a T-depth of $4n \log_2(\log_2(n/\varepsilon)) + O(n)$ with a T-count of $20n \log_2(n/\varepsilon) + O(n)$. Namely, we achieve substantial reductions in non-Clifford costs for QFT implementation, offering a cost-efficient approach to executing large-scale quantum algorithms within fault-tolerant quantum computing frameworks.

## II. OVERVIEW OF THE AQFT CIRCUITS AND THEIR ADVANTAGES

In this paper, we introduce the construction of two novel $n$-qubit AQFT circuits based on the Clifford + T gate library. AQFT Circuit 1 is optimized to minimize the T-count, whereas AQFT Circuit 2 focuses on reducing the T-depth. These AQFT circuits approximate the QFT with an error of $O(\varepsilon)$. Fig. 1(a) and Fig. 1(b) present the schematic diagrams for AQFT Circuit 1 and AQFT Circuit 2, respectively.

Similar to the AQFT circuit described in Ref. [44], the construction of our AQFT circuits involves creating $Z^\theta$ gate (See Eq. (1)) layers, each of which performs an inverse PGT, and implementing them via use of quantum adders.

$$Z^\theta = \begin{pmatrix} 1 & 0 \\ 0 & e^{i\pi\theta} \end{pmatrix}. \quad (1)$$

In order to facilitate this process, a special quantum state $|\psi_{b+1}\rangle \equiv \frac{1}{\sqrt{2^{b+1}}} \sum_{k=0}^{2^{b+1}-1} e^{2\pi i k/2^{b+1}} |k\rangle$ needs to be prepared, where $b$ is chosen as $\lceil \log_2(n/\varepsilon) \rceil$. We employ "repeat until success" circuits described in Ref. [48] to prepare the special quantum state $|\psi_{b+1}\rangle$. A detailed explanation on the inverse PGT and its execution through a quantum adder is provided in Appendix A.

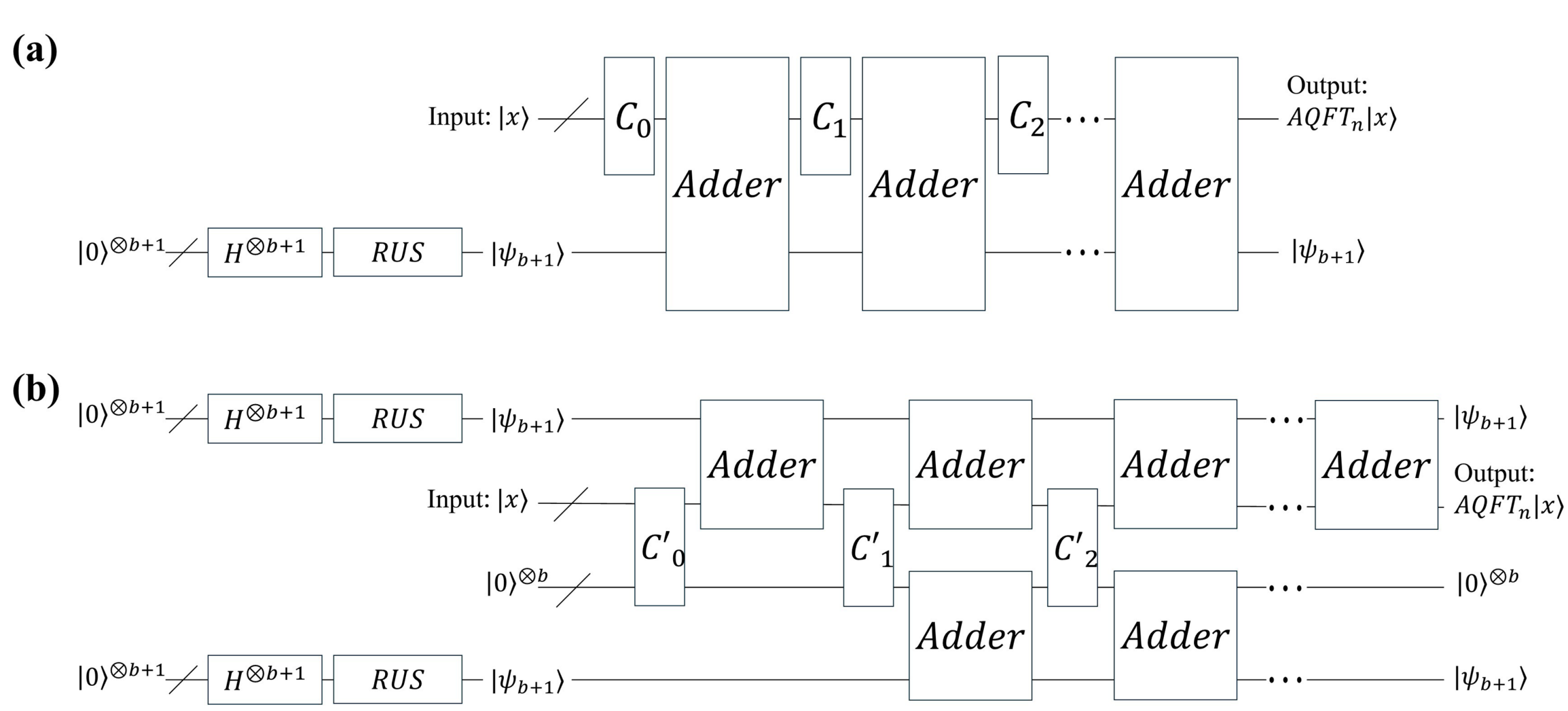

**FIG. 1**. Schematic diagram of AQFT circuits presented in this paper. In these circuits, we construct inverse PGTs, and implement them using quantum adders. The boxes labeled *RUS* represent "repeat until success" circuits from Ref. [48]. Each *RUS* circuit is used to prepare a special quantum state $|\psi_{b+1}\rangle \equiv \frac{1}{\sqrt{2^{b+1}}} \sum_{k=0}^{2^{b+1}-1} e^{2\pi i k/2^{b+1}} |k\rangle$, where $b = \lceil \log_2(n/\varepsilon) \rceil$. The special states are required to perform inverse PGTs using quantum adders. (a) AQFT Circuit 1: designed to reduce T-count. All circuits labeled $C_i$ consist of Clifford gates. (b) AQFT Circuit 2: designed to reduce T-depth. Each circuit labeled $C_i'$ consist of Clifford gates with a maximum of two T gates. Note that the quantum adders are paired and parallelized.

Before presenting detailed circuit designs and results, we provide a rough overview of the advantages of our AQFT circuits. AQFT Circuit 1 requires $\sim n$ count of $b$-qubit quantum adders to perform $\sim n$ inverse PGTs. Constructing these quantum adders involves a T-count of $\sim 4nb$. A key advantage of AQFT Circuit 1 is that the construction of inverse PGTs does not require additional non-Clifford gates, such as Toffoli gates (Note that each "$C_i$" box in Fig. 1(a) consists solely of Clifford gates.). This omission of non-Clifford gates results in a reduction of the T-count from $\sim 8nb$ to $\sim 4nb$.

AQFT Circuit 2 reduces T-depth from $O(nb)$ to $O(n\log(b))$ compared to AQFT Circuit 1 by employing the logarithmic-depth quantum adder in Ref. [47]. The inverse PGTs are paired and parallelized, which requires only additional $n + O(1)$ T gates, further diminishing the T-depth. These additional T gates are included within the $C'_i$ boxes illustrated in Fig. 1(b). This parallelization decreases the T-depth of the AQFT circuit from $\sim 8n\log_2(b)$ to $\sim 4n\log_2(b)$.

In the subsequent sections, for each AQFT circuit, we present the circuit design process, resource estimation focused on T-count and T-depth, and approximation error analysis.

## III. AQFT CIRCUIT 1: OPTIMIZED FOR T-COUNT

### A. Circuit design

The AQFT Circuit 1 is constructed through the following four steps:

*(Step 1: QFT subcircuit decomposition)* We begin by decomposing the subcircuits of the standard QFT circuit described in Ref. [49]. The initial structure of these subcircuits is illustrated on the left-hand side of Fig. 2. The right-hand side of Fig. 2 shows the decomposed structure of the QFT subcircuits. A detailed explanation of this decomposition process is provided in Appendix B. In the decomposed subcircuit, multiple $Z^{\theta}$ gate layers are present. However, in the subsequent step, we will consolidate these layers. Specifically, the $Z^{\theta}$ gates within the red boxes from all the decomposed QFT subcircuits in Fig. 2 will be combined into a single $Z^{\theta}$ gate layer and placed at the very front of the QFT circuit. Similarly, the $Z^{\theta}$ gates in the purple boxes from all the decomposed QFT subcircuits in Fig. 2 will be merged into a $Z^{\theta}$ gate layer at the very end of the QFT circuit.

*(Step 2: Constructing QFT circuit)* We combine all the decomposed QFT subcircuits to construct a full QFT circuit. Next, all the $Z^{\theta}$ gates in the red boxes from the decomposed subcircuits in Fig. 2 are gathered at the very front of the QFT circuit. This reordering is possible because both the circuits in the green and red boxes of Fig. 2 have diagonal matrix representations and therefore commute with each other. Similarly, we gather the $Z^{\theta}$ gates in the purple boxes from the decomposed subcircuits in Fig. 2 at the very end of the QFT circuit. After this reorganization, the $Z^{\theta}$ gates at the very front and end of the QFT circuit have the form of $Z^{2^{k-1}-1/2^k}$, where $k \in \{1,2,\ldots,n\}$. We divide each $Z^{2^{k-1}-1/2^k}$ gate into $Z^{1/2}$ and $Z^{-1/2^k}$ gates. Note that the $Z^{1/2}$ gate is a Clifford gate.

*(Step 3: Approximation)* We remove all the $Z^{-1/2^k}$ gates with $k > b$ from the QFT circuit (See Fig. 3).

*(Step 4: Inverse PGT implementation using quantum adder)* In Fig. 3, each $Z^{\theta}$ gate layer in blue boxes performs an inverse PGT with help the of a $Z^{-1}$ gate. The effect of inserting the $Z^{-1}$ gates can be canceled by inserting $Z$ gates. We implement each inverse PGT using a quantum adder from Ref. [46].

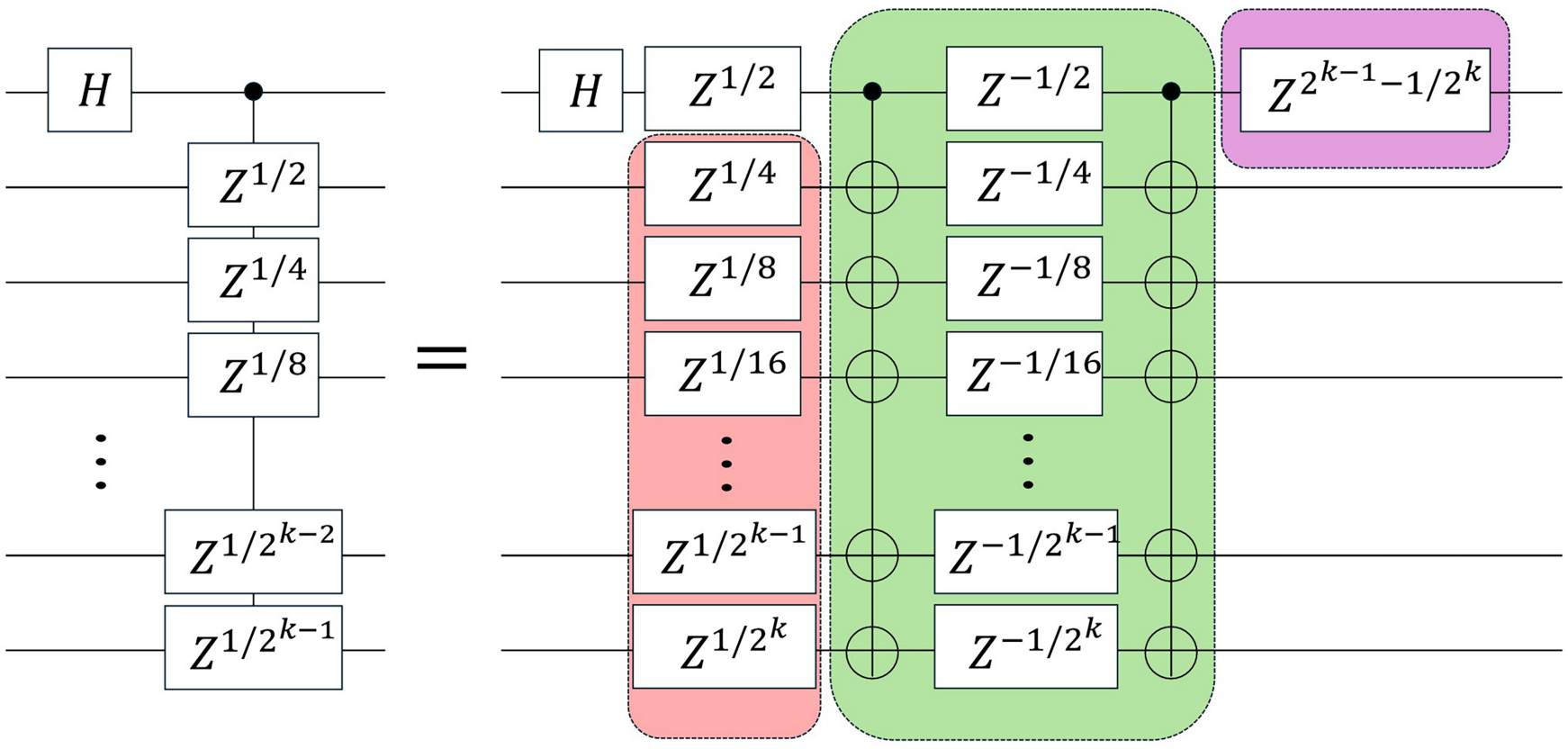

**FIG. 2**. Subcircuit decomposition for AQFT Circuit 1. The left-hand side illustrates a subcircuit of the standard QFT circuit described in Ref. [49]. This subcircuit is decomposed into structure shown on the right-hand side. The $Z^{\theta}$ gates in the red boxes from all the decomposed QFT subcircuits will be gathered into a single $Z^{\theta}$ gate layer at the very front of the QFT circuit. This reordering is possible because the circuit in the green box has a diagonal matrix representation. The $Z^{\theta}$ gate in the purple box will be gathered into a single $Z^{\theta}$ gate layer at the very end of the QFT circuit. Note that the $Z^{\theta}$ gate layer in the green box is constructed without using Toffoli gates. This results in T-count reduction.

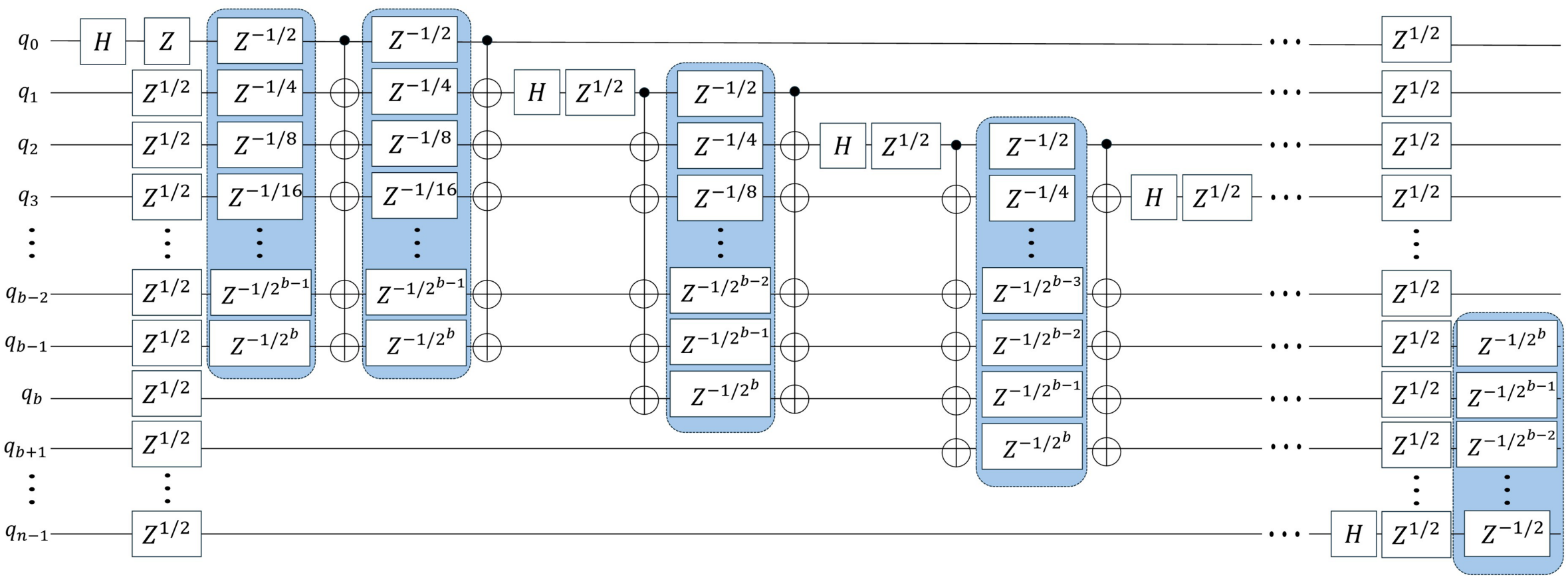

**FIG. 3**. AQFT Circuit 1: $n$-qubit AQFT with an error of $O(\varepsilon)$, where $b = \lceil \log_2(n/\varepsilon) \rceil$. This circuit is designed to optimize T-count. Each $Z^{\theta}$ gate layer in the blue boxes performs an inverse PGT with the help of a $Z^{-1}$ gate. The effect of inserting $Z^{-1}$ gates can be canceled by inserting $Z$ gates. Note that this circuit contains $(n + 1)$ inverse PGTs and these are constructed without using additional non-Clifford gates. We implement each inverse PGT using a quantum adder from Ref. [46].

### B. T-count and T-depth estimation

Among the inverse PGTs in Fig. 3, the 3-qubit inverse PGT needs not be replaced by 3-qubit quantum adder, because it only requires a single T gate. Considering this, the T gates required to construct AQFT Circuit 1 are found in four parts:

(1) $(n-b+3)$ count of $(b+1)$-qubit quantum adder,

(2) One for each of $k$-qubit quantum adder, where $k \in \{4, 5, \dots, b\}$,

(3) One T gate in the 3-qubit inverse PGT,

(4) T gates for preparing the special state $|\psi_{b+1}\rangle$.

The $t$-qubit quantum adder in Ref. [46] requires $(4t-4)$ T-count and $t$ T-depth. With the quantum adder, the T-count and T-depth in part (1) are $4nb-4b^2+12b$ and $nb+n-b^2+2b+3$, respectively. The T-count and T-depth in part (2) are $\sum_{k=4}^{b}(4k-4)=2b^2-2b-12$ and $\sum_{k=4}^{b} k = b^2/2+b/2-6$, respectively. For part (4), $(b-2)$ $Z^{\theta}$ gates have to be synthesized to error $\varepsilon/b$ for each gate (See Appendix A). The value $\varepsilon/b$ is chosen so that AQFT Circuit 1 has an approximation error of $O(\varepsilon)$. In the gate synthesis, we use $RUS$ circuits from Ref. [48], and the expectation value to synthesize a $Z^{\theta}$ gate with a Fourier angle to error $\varepsilon'$ is $1.08\log_2(1/\varepsilon')+17.5$ [50]. Therefore, the total T-count for AQFT Circuit 1 is

$$\begin{aligned} &4nb-2b^2+b\{1.08\log_2(b/\varepsilon)+27.5\}-2.16\log_2(b/\varepsilon)+O(1) \\ &= 4nb-O(b^2), \end{aligned} \tag{2}$$

and the total T-depth for AQFT Circuit 1 is

$$\begin{aligned} &nb+n-b^2/2+5b/2+1.08\log_2(b/\varepsilon)+O(1) \\ &= nb+O(n). \end{aligned} \tag{3}$$

### C. Approximation error analysis

The definition of approximation error can be found in Appendix C. We demonstrate that AQFT Circuit 1 has an error of $O(\varepsilon)$ when implemented in replacement of a QFT circuit. The approximation error in AQFT Circuit 1 arises from two sources:

(1) Removing all the $Z^{-1/2^k}$ gates with $k>b$,

(2) Approximating $Z^{\theta}$ gates to prepare the special state $|\psi_{b+1}\rangle$ using $RUS$ circuits.

We call the error from part (1) $\varepsilon_1^{(1)}$ and the error from part (2) $\varepsilon_1^{(2)}$.

The error for removing a $Z^{\theta}$ gate is

$$\max_{|\psi\rangle} \| (I - Z^{\theta}) |\psi\rangle \| \tag{4}$$

and this can be rewritten as

$$\sqrt{2(1-\cos\theta\pi)} = 2|\sin(\pi\theta/2)| < |\pi\theta|, \tag{5}$$

since

$$(I - Z^{\theta})^{\dagger} (I - Z^{\theta}) = \begin{pmatrix} 0 & 0 \\ 0 & 2(1-\cos\theta\pi) \end{pmatrix}. \tag{6}$$

Therefore, the error $\varepsilon_1^{(1)}$ is bounded as follows:

$$\varepsilon_1^{(1)} \leq (n-b-2) \cdot \sum_{j=b+1}^{n} \frac{\pi}{2^j} < n2^{-b}\pi < \varepsilon\pi. \tag{7}$$

Note that $b$ is chosen as $\lceil \log_2(n/\varepsilon) \rceil$, thus we have $n2^{-b} \leq \varepsilon$.

When synthesizing $Z^{\theta}$ gates to prepare the special state $|\psi_{b+1}\rangle$, each $Z^{\theta}$ gate is approximated to the error $\varepsilon/b$, resulting in $\varepsilon_1^{(2)} \leq (b-2) \cdot \frac{\varepsilon}{b} < \varepsilon$. Consequently, the total error implementing AQFT Circuit 1 instead of the QFT circuit is $O(\varepsilon)$ because

$$\varepsilon_1^{(1)} + \varepsilon_1^{(2)} < (1+\pi)\varepsilon. \tag{8}$$

## IV. AQFT CIRCUIT 2: OPTIMIZED FOR T-DEPTH

### A. Circuit design

The construction of AQFT Circuit 2 involves the following four steps:

*(Step 1: QFT subcircuit decomposition)* We begin by pairing the subcircuits on the left-hand side of Fig. 2, as shown on the left-hand side of Fig. 4. These paired subcircuits are then decomposed into the structure illustrated on the right-hand side of Fig. 4. For the decomposition of a $k$-qubit subcircuit, $(k-1)$ ancilla qubits, initially in the state $|0\rangle^{\otimes(k-1)}$, are required. A detailed demonstration of this decomposition is provided in Appendix B. In the decomposed structure, note that two $Z^{\theta}$ gate layers are paired and parallelized within the green box in Fig. 4. Similar to AQFT Circuit 1, in the next step, we will combine the $Z^{\theta}$ gates in the red boxes from all the decomposed QFT subcircuits in Fig. 4 into a single $Z^{\theta}$ gate layer at the very front of the QFT circuit. Additionally, the $Z^{\theta}$ gates in the purple boxes from all the decomposed QFT subcircuits in Fig. 4 will be gathered into a single $Z^{\theta}$ gate layer at the very end of the QFT circuit.

*(Step 2: Constructing QFT circuit)* We combine all the decomposed QFT subcircuits to construct a full QFT circuit. We modify the circuit using a method similar to that of AQFT

Circuit 1. All the $Z^{\theta}$ gates in the red boxes from the decomposed subcircuits in Fig. 4 are gathered at the very front of the QFT circuit, and the $Z^{\theta}$ gate in the purple boxes from the decomposed subcircuits in Fig. 4 are gathered at the very end of the QFT circuit. Then, the $Z^{\theta}$ gates at the very front and end of the QFT circuit take the form of $Z^{2^{k-1}-1/2^{k}}$, where $k \in \{1,2,\ldots,n\}$. We divide each $Z^{2^{k-1}-1/2^{k}}$ gate into $Z^{1/2}$ and $Z^{-1/2^{k}}$ gates.

*(Step 3: Approximation)* We remove all the $Z^{-1/2^{k}}$ gates with $k > b$ from the QFT circuit (See Fig. 5). In the approximation process, we no longer need $(n-1)$ ancilla qubits; instead, we only need $b$ ancilla qubits.

*(Step 4: Inverse PGT implementation using quantum adder)* In Fig. 5, each $Z^{\theta}$ gate layer in blue boxes performs an inverse PGT with help the of a $Z^{-1}$ gate. The effect of inserting the $Z^{-1}$ gates can be canceled by inserting $Z$ gates. We implement each inverse PGT using a quantum adder (In-FT-QCLA1) from Ref. [47].

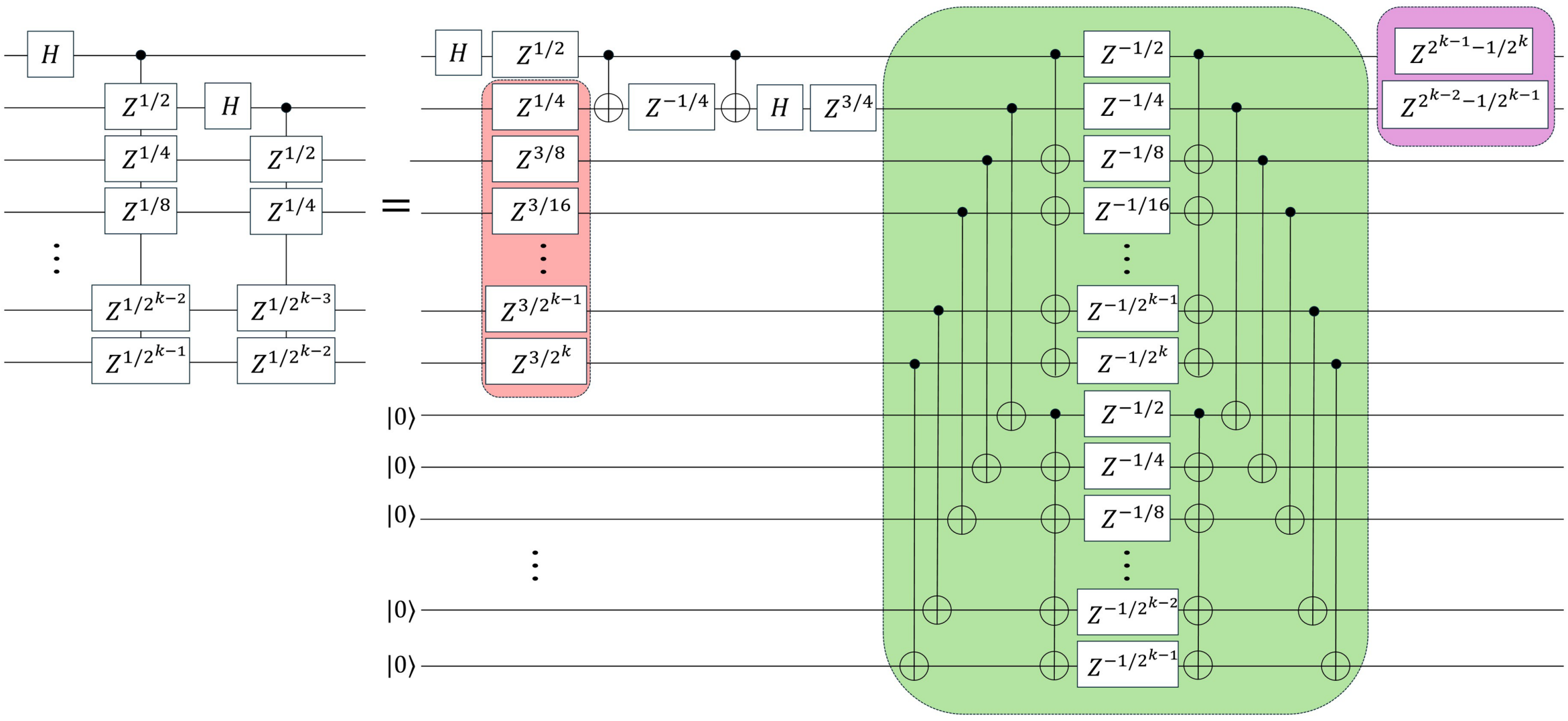

**FIG. 4**. Subcircuit decomposition for AQFT Circuit 2. The $Z^{\theta}$ gates in the red boxes from all the QFT subcircuits will be gathered into a single $Z^{\theta}$ gate layer at the very front of the QFT circuit. This reordering is possible because the circuits in the red and green boxes have diagonal matrix representations. The $Z^{\theta}$ gates in the purple box will be gathered into a single $Z^{\theta}$ gate layer at the very end of the QFT circuit. Note that two $Z^{\theta}$ gate layers are paired and parallelized in the green box. This results in T-depth reduction.

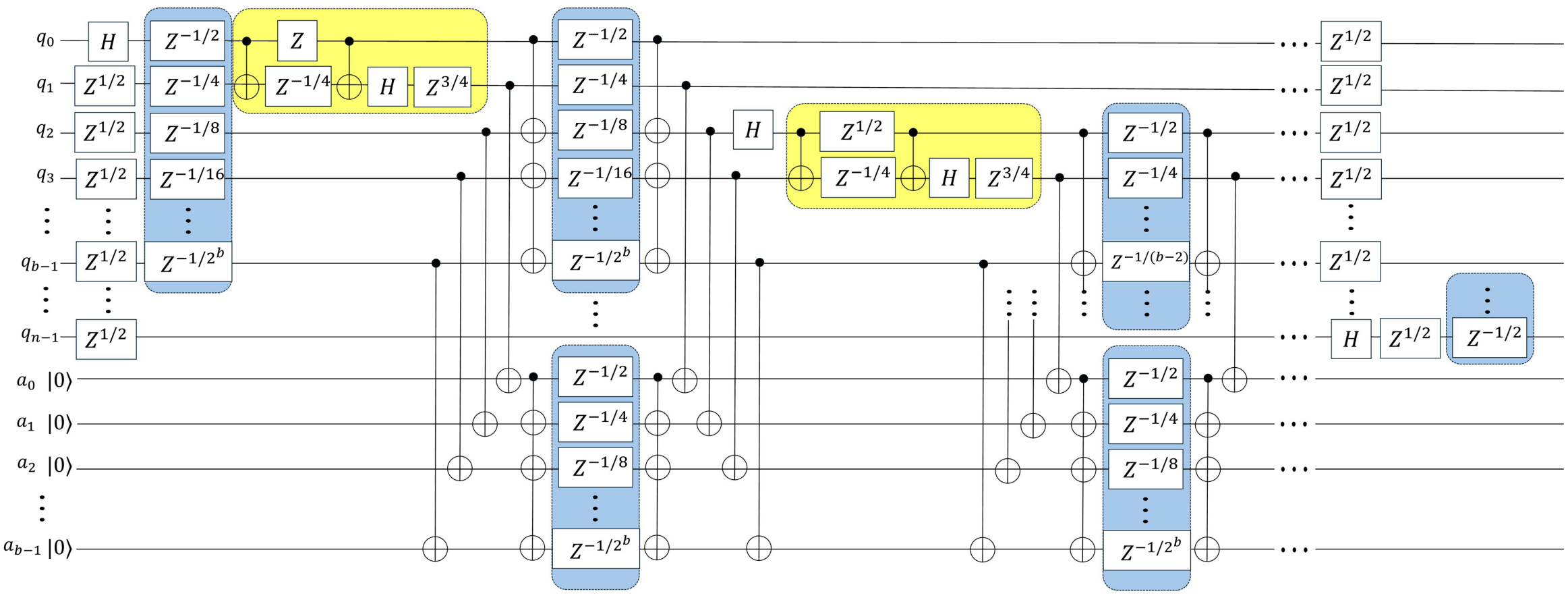

**FIG. 5**. AQFT Circuit 2: $n$-qubit AQFT with an error of $O(\varepsilon)$ optimized for T-depth, where $b = \lceil \log_2(n/\varepsilon) \rceil$. This circuit employs $b$ ancilla qubits each of which is labeled as $a_i$ and initially in the state $|0\rangle$. Each $Z^{\theta}$ gate layer in the blue boxes performs an inverse PGT with the help of a $Z^{-1}$ gate. The effect of inserting $Z^{-1}$ gates can be canceled by inserting $Z$ gates. Note that this circuit contains $(n+1)$ inverse PGTs, and each yellow box contains only two T gates. We implement each inverse PGT using a logarithmic-depth quantum adder (In-FT-QCLA1) from Ref. [47].

## B. T-count and T-depth estimation

Among the inverse PGTs in Fig. 5, the 3-qubit inverse PGT does not need to be replaced by 3-qubit quantum adder, because it only requires a single T gate. Considering this, the T gates required to construct AQFT Circuit 2 are found in five parts:

(1) $(n-b+3)$ count of $(b+1)$-qubit quantum adder,

(2) One for each of $k$-qubit quantum adder, where $k \in \{4, 5, \dots, b\}$,

(3) One T gate in 3-qubit inverse PGT,

(4) $(n-1)$ T gates in yellow boxes in Fig. 5,

(5) T gates for preparing two of the special states $|\psi_{b+1}\rangle$.

The $t$-qubit quantum adder (In-FT-QCLA1) requires $20t - 8w(t) - 8w(t-1) - 4\lfloor \log_2(t) \rfloor - 4\lfloor \log_2(t-1) \rfloor - 8$ T-count, where $w(t) = t - \sum_{i=1}^{\infty} \lfloor t/2^i \rfloor$ represents the number of ones in the binary expansion of $t$, and $2\lfloor \log_2(t) \rfloor + 2\lfloor \log_2(t-1) \rfloor + 2\lfloor \log_2(t/3) \rfloor + 2\lfloor \log_2((t-1)/3) \rfloor + 28$ T-depth [47]. For readability, we use the following notations:

$$TC_{adder}(t) = 20t - 8w(t) - 8w(t-1) - 4\lfloor \log_2(t) \rfloor - 4\lfloor \log_2(t-1) \rfloor - 8,$$

$$TD_{adder}(t) = 2\lfloor \log_2(t) \rfloor + 2\lfloor \log_2(t-1) \rfloor + 2\lfloor \log_2(t/3) \rfloor + 2\lfloor \log_2((t-1)/3) \rfloor + 28 . \quad (9)$$

In the following resource estimation, for the sake of convenience in calculation, we assume that $n$ is even and $b$ is odd. Even without this assumption, however, the difference in resource calculation is only $O(1)$. Then, the T-count and T-depth in part (1) are $(n-b+3)\cdot TC_{adder}(b)$ and $\{2+\frac{n-b+1}{2}\}\cdot TD_{adder}(b)$, respectively. The T-count and T-depth in part (2) is $\sum_{i=4}^{b-1} TC_{adder}(i)$ and $\sum_{i=2}^{(b-1)/2} TD_{adder}(2i)$, respectively. For T gates in part (5), $2(b-2)$ $Z^{\theta}$ gates have to be synthesized to error $\varepsilon/2b$ for each gate (See Appendix A). The value $\varepsilon/2b$ is chosen to ensure that AQFT Circuit 2 has the approximation error, $O(\varepsilon)$. Therefore, the total T-count for AQFT Circuit 2 is

$$\begin{aligned}(n-b+3)\cdot TC_{adder}(b)+\sum_{i=4}^{b-1} TC_{adder}(i)\\ +n+2b(1.08\log_2(b/\varepsilon)+17.5)-4.32\log_2(b/\varepsilon)+O(1)\\ =20nb+O(n),\end{aligned} \tag{10}$$

and the total T-depth for AQFT Circuit 2 is

$$\begin{aligned}\left\{\frac{n-b+5}{2}\right\}\cdot TD_{adder}(b)+\sum_{i=2}^{\frac{b-1}{2}} TD_{adder}(2i)\\ +n+1.08\log_2(b/\varepsilon)+O(1)\\ =4n\log_2(b)+O(n).\end{aligned} \tag{11}$$

**C. Approximation error analysis**

We prove that AQFT Circuit 2 has an error of $O(\varepsilon)$ when implemented in place of a QFT circuit. The sources of the approximation error in AQFT Circuit 2 are identified in two parts:

(1) Removing all the $Z^{-1/2^k}$ gates with $k>b$,

(2) Approximating $Z^{\theta}$ gates to prepare two of the special states $|\psi_{b+1}\rangle$ using $RUS$ circuits.

We call the error from part (1) $\varepsilon_2^{(1)}$ and the error from part (2) $\varepsilon_2^{(2)}$.

The error $\varepsilon_2^{(1)}$ is bounded in the same way as $\varepsilon_1^{(1)}$. When synthesizing $Z^{\theta}$ gates to prepare two of the special states $|\psi_{b+1}\rangle$ for AQFT Circuit 2, each $Z^{\theta}$ gate is approximated to the error $\varepsilon/2b$, resulting in $\varepsilon_2^{(2)}\leq 2(b-2)\cdot\frac{\varepsilon}{2b}<\varepsilon$. Consequently, the total error implementing AQFT Circuit 2 instead of the QFT circuit is $O(\varepsilon)$.

## V. DISCUSSION

In summary, we introduced two novel $n$-qubit AQFT circuits: AQFT Circuit 1 for T-count optimization and AQFT Circuit 2 for T-depth optimization. AQFT Circuit 1 is designed to minimize T-count and achieves a significant reduction with a T-count of $4n\log_2(n/\varepsilon) - O(\log^2(n/\varepsilon))$ which is roughly half of the best-known result in Ref. [44]. This reduction is accomplished by constructing inverse PGT circuits without using Toffoli gates and by replacing them with quantum adders from Ref. [46]. On the other hand, AQFT Circuit 2 is designed to minimize T-depth, reducing it from $O(n\log(n/\varepsilon))$ to $O(n\log(\log(n/\varepsilon)))$. This is achieved by employing logarithmic-depth quantum adders from Ref. [47]. We further reduce the T-depth by pairing and parallelizing inverse PGTs using only $n + O(1)$ additional T gates. These results are summarized in Table I.

**TABLE I.** Comparison of T-count and T-depth for various $n$-qubit AQFT implementations with an approximation error of $O(\varepsilon)$. In the table, we use notation $b$, $TC_{adder}(t)$, and $TD_{adder}(t)$ for readability, where $b = \lceil\log_2(n/\varepsilon)\rceil$, $TC_{adder}(t) = 20t - 8w(t) - 8w(t-1) - 4\lfloor\log_2(t)\rfloor - 4\lfloor\log_2(t-1)\rfloor - 8$, and $TD_{adder}(t) = 2\lfloor\log_2(t)\rfloor + 2\lfloor\log_2(t-1)\rfloor + 2\lfloor\log_2(t/3)\rfloor + 2\lfloor\log_2((t-1)/3)\rfloor + 28$. $w(t) = t - \sum_{i=1}^{\infty}\lfloor t/2^i\rfloor$ represents the number of ones in the binary expansion of $t$. For the sake of convenience in calculation, we assume that $n$ is even and $b$ is odd. Even without this assumption, the difference in resource calculation is only $O(1)$.

| Source | T-count | T-depth |
| --- | --- | --- |
| Ref. [44] | $8nb - 4b^2$<br>$+25.5b + 2.16b\log_2(b/\varepsilon)$<br>$-2.16\log_2(b/\varepsilon) + O(1)$<br>$= 8nb - O(b^2)$ | $nb + 5n - b^2/2 + b/2$<br>$+1.08\log_2(b/\varepsilon) + O(1)$<br>$= nb + O(n)$ |
| AQFT Circuit 1 | $4nb - 2b^2$<br>$+b\{1.08\log_2(b/\varepsilon) + 27.5\}$<br>$-2.16\log_2(b/\varepsilon) + O(1)$<br>$= 4nb - O(b^2)$ | $nb + n - b^2/2 + 5b/2$<br>$+1.08\log_2(b/\varepsilon) + O(1)$<br>$= nb + O(n)$ |
| AQFT Circuit 2 | $(n - b + 3) \cdot TC_{adder}(b)$<br>$+\sum_{i=4}^{b-1} TC_{adder}(i)$<br>$+n + 2b(1.08\log_2(b/\varepsilon) + 17.5)$<br>$-4.32\log_2(b/\varepsilon) + O(1)$<br>$= 20nb + O(n)$ | $\{(n - b + 5)/2\} \cdot TD(b)$<br>$+\sum_{i=2}^{\frac{b-1}{2}} TD_{adder}(2i) + n$<br>$+1.08\log_2(b/\varepsilon) + O(1)$<br>$= 4n\log_2 b + O(n)$ |

Considering that T gates are the main obstacle for implementing quantum algorithms fault-tolerantly, and that QFT is ubiquitous in quantum algorithms, we expect that our results can accelerate the use of QFT in large-scale quantum algorithms such as Shor's factoring algorithm [2] and the HHL algorithm [3]. The results of this study underscore the potential to significantly enhance the execution of large-scale quantum algorithms, paving the way for improved efficiency.

The main limitation of our study is that the T-count and T-depth of the AQFT circuits are not simultaneously optimized. However, our findings demonstrate significant improvements in either T-count or T-depth separately, providing a foundation for further optimization. Additionally, we did not provide any theoretical bound of T-count or T-depth for AQFT implementation.

Further research should focus on optimizing both the T-count and T-depth simultaneously, especially to achieve both a T-depth of $O(nb)$ and a T-count of $4nb - O(b^2)$. One promising avenue is the development of new quantum adders with reduced T-count and T-depth. Additionally, we hypothesize that AQFT Circuit 1 has the smallest T-count among AQFT implementation methods that create PGT circuits and execute them using quantum adders. Further research is required to prove that AQFT Circuit 1 indeed has the smallest T-count among such AQFT implementation methods.

## ACKNOWLEDGMENTS

The authors thank Sangkyu Baek for his valuable comments. This work was supported by Korea National Research Foundation (NRF) grant No. NRF-2023R1A2C1003570, RS-2023-00225385, RS-2024-00422330, AFOSR grant FA2386-21-1-0089, AFOSR grant FA2386-22-1-4052, and Amazon Web Services. This work was also supported by the National Quantum Laboratory at Maryland (QLab).

## APPENDIX A: PHASE GRADIENT TRANSFORMATION

The PGT on a $t$-qubit system and its inverse are defined as follows:

$$\begin{aligned} PGT_t|x\rangle &= e^{2\pi i x/2^t}\,|x\rangle, \\ PGT_t^{-1}|x\rangle &= e^{-2\pi i x/2^t}|x\rangle, \end{aligned} \tag{A1}$$

where $|x\rangle$ is a $t$-qubit computational basis state.

In this paper, we utilize inverse PGTs. Here, we present two methods for implementing inverse PGT. The first method employs a layer of $Z^{\theta}$ gates, where $\theta \in \{-1, -1/2, -1/4, \dots, -1/2^{t-1}\}$. Fig. 6 illustrates the circuit of $Z^{\theta}$ gates used to implement inverse PGT.

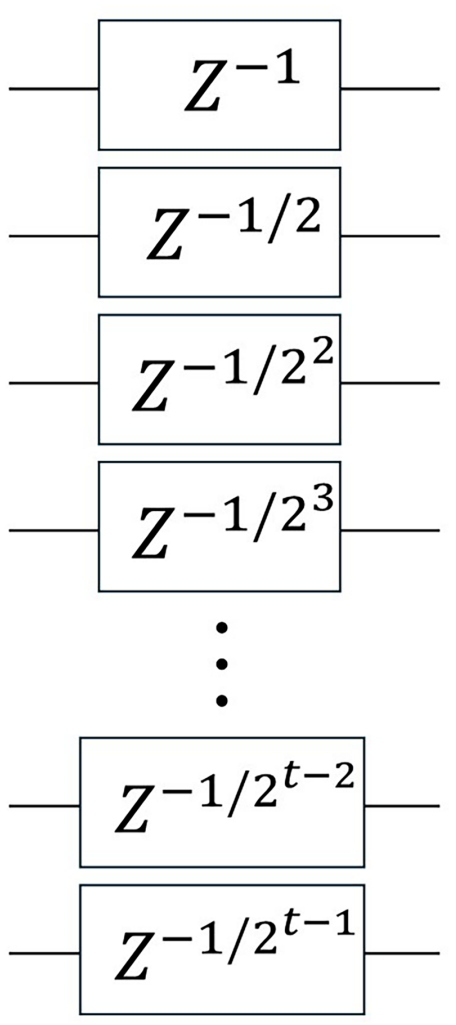

**FIG. 6**. Circuit diagram using $Z^{\theta}$ gates for implementing $t$-qubit inverse PGT.

The alternative involves quantum addition, which is originally reported in Ref. [45]. Adding the state $|x\rangle$ with a special state $|\psi_t\rangle = \frac{1}{\sqrt{2^t}} \sum_{j=0}^{2^t-1} e^{2\pi i j/2^t} |j\rangle$ can perform the inverse PGT on the state $|x\rangle$:

$$
\begin{aligned}
Adder|x\rangle|\psi_t\rangle &= \frac{1}{\sqrt{2^t}} \sum_{j=0}^{2^t-1} e^{\frac{2\pi i j}{2^t}} |x\rangle |(x+j) \bmod 2^t\rangle \\
&= \frac{1}{\sqrt{2^t}} \sum_{j=0}^{2^t-1} e^{-\frac{2\pi i x}{2^t}} e^{\frac{2\pi i (x+j)}{2^t}} |x\rangle |(x+j) \bmod 2^t\rangle \\
&= e^{-\frac{2\pi i k}{2^t}} |x\rangle \frac{1}{\sqrt{2^t}} \sum_{j=0}^{2^t-1} e^{\frac{2\pi i (x+j)}{2^t}} |(x+j) \bmod 2^t\rangle \\
&= e^{-\frac{2\pi i x}{2^t}} |x\rangle |\psi_t\rangle \\
&= PGT_t^{-1} |x\rangle |\psi_t\rangle.
\end{aligned}
\tag{A2}
$$

The state $|\psi_t\rangle$ is prepared by applying $t$ number of $H$ gates and a $Z^{\theta}$ gates layer to the state $|0\rangle^{\otimes t}$. We use the $RUS$ circuit from Ref. [48] to construct the $Z^{\theta}$ gates. Fig. 7 illustrates the circuit of implementing inverse PGT using a quantum adder.

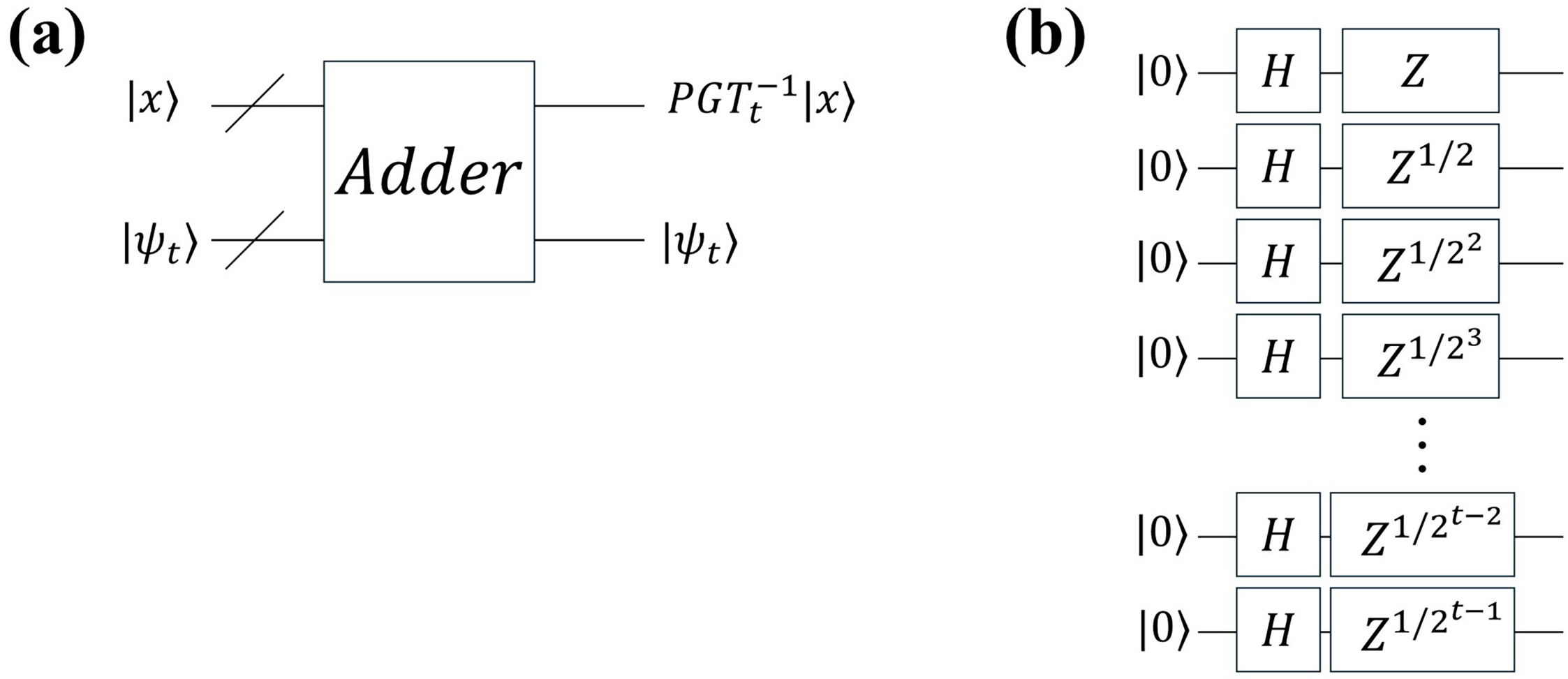

**FIG. 7**. Circuit diagram for implementing inverse PGT using a quantum adder. (a) Schematic diagram. (b) Circuit for preparing the state $|\psi_t\rangle$. We use the $RUS$ circuit from Ref. [48] for constructing the $Z^{\theta}$ gates.

This quantum addition-based method has two notable features. First, the state $|\psi_t\rangle$ is preserved after inverse PGT implementation, meaning $|\psi_t\rangle$ is reusable. Second, if we have the state $|\psi_t\rangle$ and wish to implement the inverse PGT on a $t'$-qubit system, where $t' < t$, the state $|\psi_{t'}\rangle$ does not need to be prepared separately; the state of the partial qubits in $|\psi_t\rangle$ is simply $|\psi_{t'}\rangle$.

## APPENDIX B: SUBCIRCUIT CONSTRUCTIONS

First, we show that both sides of Fig. 2 evolve in the same way. For the proof, we ignore the common $H$ gate on the first qubit and assume that the initial state of each side is in the computational basis state $|x\rangle = |x_0 x_1 \dots x_{k-1}\rangle$, where $x_i$ is 0 or 1. The state of the left-hand side circuit without $H$ gate in Fig. 2 evolves into the state

$$e^{\pi i x_0 \left(\frac{x_1}{2} + \frac{x_2}{2^2} + \frac{x_3}{2^3} + \cdots \frac{x_{k-1}}{2^{k-1}}\right)} |x_0 x_1 \dots x_{k-1}\rangle, \tag{B1}$$

and the state of the right-hand side circuit without $H$ gate in Fig. 2 evolves into the state

$$e^{\pi i \left(\frac{x_0}{2} + \frac{x_1}{2^2} + \frac{x_2}{2^3} + \cdots \frac{x_{k-1}}{2^k}\right)} e^{-\pi i \left(\frac{x_0}{2} + \frac{x_0 \oplus x_1}{2^2} + \frac{x_0 \oplus x_2}{2^2} + \cdots \frac{x_0 \oplus x_{k-1}}{2^k}\right)} e^{\pi i x_0 \left(\frac{2^{k-1} - 1}{2^k}\right)} |x_0 x_1 \dots x_{k-1}\rangle. \tag{B2}$$

We show that equations (B1) and (B2) are equal by considering the cases where $x_0$ is 0 or 1. If $x_0 = 0$, both equations (B1) and (B2) reduce to $|x_0 x_1 \dots x_{k-1}\rangle$. If $x_0 = 1$, then the equation (B1) becomes

$$e^{\pi i(\frac{x_1}{2}+\frac{x_2}{2^2}+\frac{x_3}{2^3}+\cdots\frac{x_{k-1}}{2^{k-1}})}|x_0x_1\ldots x_{k-1}\rangle, \tag{B3}$$

and equation (B2) becomes

$$\begin{aligned}
&e^{\pi i\left(\frac{x_1}{2^2}+\frac{x_2}{2^3}+\cdots\frac{x_{k-1}}{2^k}\right)}e^{-\pi i\left(+\frac{1\oplus x_1}{2^2}+\frac{1\oplus x_2}{2^2}+\cdots\frac{1\oplus x_{k-1}}{2^k}\right)}e^{\pi i\left(\frac{2^{k-1}-1}{2^k}\right)}|x_0x_1\ldots x_{k-1}\rangle \\
&= e^{\pi i\left(\frac{2x_1-1}{2^2}+\frac{2x_2-1}{2^3}+\cdots\frac{2x_{k-1}-1}{2^k}\right)}e^{\pi i\left(\frac{2^{k-1}-1}{2^k}\right)}|x_0x_1\ldots x_{k-1}\rangle \\
&= e^{\pi i\left(\frac{x_1}{2}+\frac{x_2}{2^2}+\cdots\frac{x_{k-1}}{2^{k-1}}\right)}e^{-\pi i\left(\frac{1}{2^2}+\frac{1}{2^3}+\cdots\frac{1}{2^k}\right)}e^{\pi i\left(\frac{2^{k-1}-1}{2^k}\right)}|x_0x_1\ldots x_{k-1}\rangle \\
&= e^{\pi i(\frac{x_1}{2}+\frac{x_2}{2^2}+\frac{x_3}{2^3}+\cdots\frac{x_{k-1}}{2^{k-1}})}|x_0x_1\ldots x_{k-1}\rangle.
\end{aligned} \tag{B4}$$

Therefore, equation (B1) and (B2) are the same. Note that $y-(1\oplus y)=2y-1$, where $y$ is 0 or 1.

Next, we prove the subcircuit construction in Fig. 4. We pair the subcircuits in the standard QFT circuit [49] and decompose them using the method used in subcircuit construction for AQFT Circuit 1 (See Fig. 8).

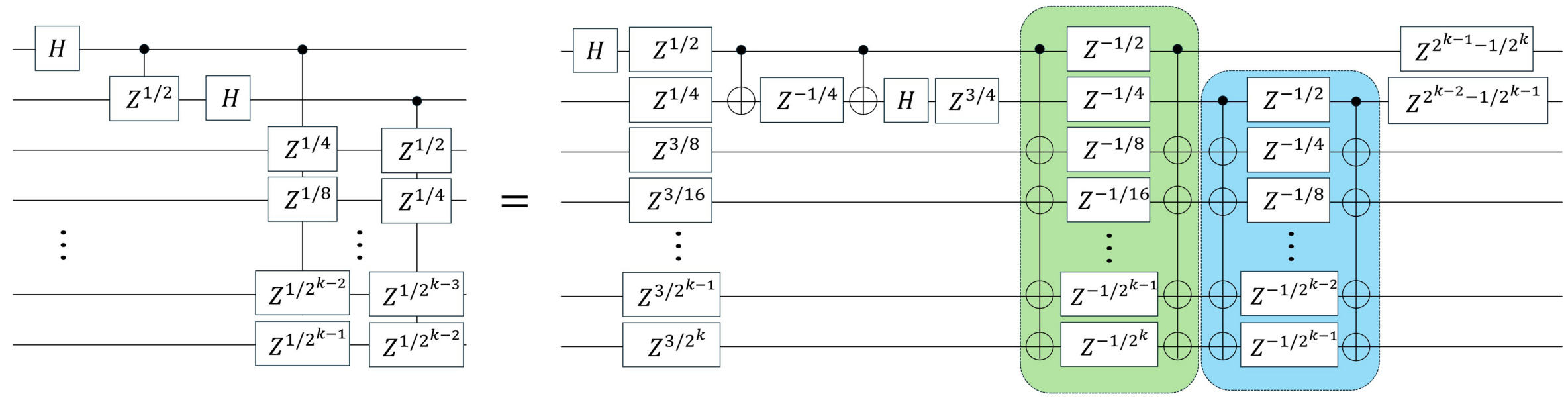

**FIG. 8**. Paired subcircuits of QFT and its decomposition.

Then, we modify the right-hand side circuit of Fig. 8 into the right-hand side circuit of Fig. 4 by using the circuit identity shown in Fig. 9. Note that each circuit in the green box and blue box in Fig. 8 has a diagonal matrix representation, thus the circuit identity in Fig. 9 is applicable.

The circuit identity in Fig. 9 can be proved trivially using basic matrix operations. Assume both $D_1$ and $D_2$ in Fig. 9 are operators on $t$-qubit systems and their matrix representations are $diag(d_{10},d_{11},\ldots,d_{1(2^t-1)})$ and $diag(d_{20},d_{21},\ldots,d_{2(2^t-1)})$, respectively. Let the state before performing $D_1$ and $D_2$ be in the computational basis state $|x\rangle$, where $x\in\{0,1,\ldots,2^t-1\}$. Then, after performing the circuit on either the left-hand side or right-hand side of Fig. 9, the state evolves into the same state $d_{1x}d_{2x}|x\rangle$. Therefore, the circuits on the left-hand side and right-hand side of Fig. 9 perform the same operation.

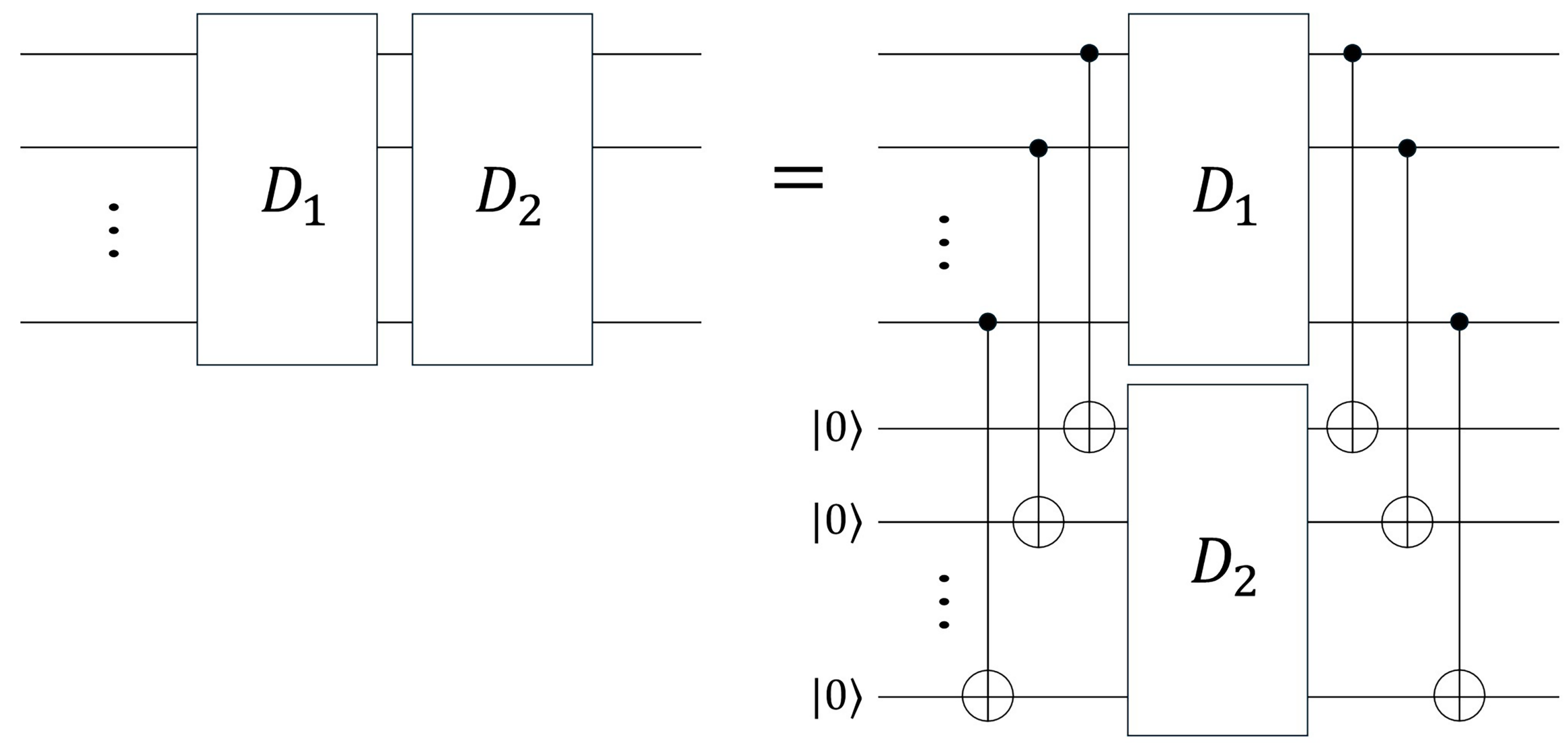

**FIG. 9**. Circuit identity to parallelize circuits each of which has diagonal matrix representation. The boxes labeled $D_1$ and $D_2$ are circuits with diagonal matrix representation.

## APPENDIX C: APPROXIMATION ERROR

The error when implementing circuit $V$ instead of $U$ is defined by

$$E(U,V) \equiv \max_{|\psi\rangle} \| (U - V)|\psi\rangle \|, \tag{C1}$$

where the maximum is taken over all normalized quantum states $|\psi\rangle$ in the state space [49]. The accumulative error is at most the sum of individual errors given by:

$$E(U_1 U_2 \dots U_n, V_1 V_2 \dots V_n) \leq \sum_{i=1}^{n} E(U_i, V_i). \tag{C2}$$